\documentclass{revtex4}
\usepackage[utf8]{inputenc}
\usepackage[T1]{fontenc}
\usepackage{lmodern}
\usepackage{graphicx}
\usepackage[figurename=Fig.,labelfont=bf,labelsep=period]{caption}
\usepackage{subfigure}
\usepackage{amsmath}
\usepackage{amsthm}
\usepackage{newtxtext,newtxmath}

\usepackage{algorithm}
\usepackage{algorithmic}

\usepackage[colorlinks=true,citecolor=red,linkcolor=black]{hyperref}

\theoremstyle{definition} \newtheorem{defi}{Definition}

\def\deltap{\delta^+}

\def\deltap{\partial^+}

\def\sit{s_i^t}
\def\yit{y_i^t}

\def\ui{v_i}

\begin{document}

% You will need to make the title all-caps
%\title{Template for Preparing Your Submission to the American Society Of Civil Engineers (ASCE)}
\title{Contamination Source Detection in Water Distribution Networks using Belief Propagation.} %Also required!!!

\author{E. Ortega}
%\email{ernesto.ortega@fisica,uh.cu}
\affiliation{Facultad de F\'{\i}sica, Universidad de la Habana. Cuba}
\author{A. Braunstein}
\affiliation{DISAT, Politecnico di Torino, Corso duca degli Abruzzi 24, Turin, Italy
IIGM, Via Nizza 52, Turin, Italy, Collegio Carlo Alberto, Moncalieri, Italy.}
\author{A. Lage-Castellanos}
\email{ale.lage@gmail.com}
\affiliation{Facultad de F\'{\i}sica, Universidad de la Habana. Cuba}

\begin{abstract}
We present a Bayesian approach for the Contamination Source Detection problem in Water Distribution Networks. Given an observation of contaminants in one or more nodes in the network, we try to give probable explanation for it assuming that contamination is a rare event. We introduce extra variables to characterize the place and pattern of the first contamination event. Then we write down the posterior distribution for these extra variables given the observation obtained by the sensors. Our method relies on Belief Propagation for the evaluation of the marginals of this posterior distribution and the determination of the most likely origin. The method is implemented on a simplified binary forward-in-time dynamics. Simulations on data coming from the realistic simulation software EPANET on two networks show that the simplified model is nevertheless flexible enough to capture crucial information about contaminant sources.
\end{abstract}

\maketitle

% Please include an abstract:

\section{Introduction}

\begin{figure}
\centering
\includegraphics[width=0.5\textwidth]{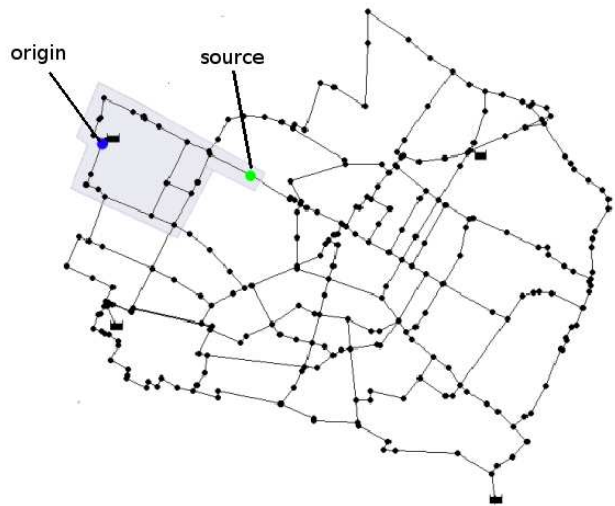}
\caption{\label{fig:modenaepanet} Graph of Modena city viewed with EPANET. Node marked in blue represent the source of the contamination and node marked in green represent the sensor position}
\end{figure}

Contamination Source Detection (CSD) is an inverse problem where, after detecting the presence of contamination in some few nodes of a water distribution network (see Fig. \ref{fig:modenaepanet} for an example), it is required to estimate the most probable origin.
It has received a lot of attention in the recent past, both for economical and safety reasons. 
%The importance of the drinking water and the decreasing existence of it in the largest cities of the world, makes it of maximum importance every effort in keep the best quality of the water destined for the consume of population. The early detection of the source of a contamination spread in a Water Distribution Network (WDN) could be essential in the prevention of future or further damage.   

The network is generally considered to be known, to some level of detail. A first challenge arises from the inverse nature of the problem, even assuming a perfect knowledge of the parameters of the model. In such cases, a trial and error scheme has been normally attempted \cite{REVERSE,laird2006mixed,laird2007real,liu2011logistic}. A further challenge comes from all the sources of stochasticity, as could be the fluctuating demand, non reliable sensors or uncertainty in the parameters defining the network. Researchers have approached this more realistic case in a probabilistic manner \cite{preis2006contamination,huang2009data,hu2015mapreduce,guan,tao2012identification,kumar2012contaminant,khancontamination,wang2011bayesian,wang2013bayesian,wang2012improving,propato,barandouzi2016probabilistic,cristo2008pollution,perelman2010bayesian}.

Both previous approaches regard as the main magnitude the time pattern of concentration of contaminants $C_i(t)$ 
in each node $i\in WDN$. It is what they read from the sensor, and what they want to infer in the source node. Many studies relies on a two steps approach, first selecting a reduced set of candidates nodes (by discarding those that can not reach the sensor, for instance) and then try some brute force trial and error approach, in which possible origins are proposed and then the corresponding pattern is optimized to match the observation. In the search for more simple approaches, we showed in \cite{paper} how to transform the CSD problem into a linear optimization one, under a strong simplification, that of considering only binary states for the process such that $C_i(t) \in \{0,1\}$. The underlying idea being that the temporal binary pattern (something like $..0,0,1,1,1,0,0,1,0,1,0,0,0..$ using some time discretization) observed at the sensor could carry enough information to locate the source.

In the present work we built upon the same simplification, that of binary clean/contamination states to approach the CSD problem from the perspective of statistical mechanics. Other statistical mechanics approaches have shown in the past to be useful in the formally similar problem of source detection in epidemic compartment models on networks\cite{lokhov2014inferring,luo2014identify,zhu2016information,Ale}. To the extent of our knowledge this approach is new in the context of contamination source detection. It should be noted that the proposed inference method can also be applied to other discrete (or comparment) propagation models on networks, such as SI, SIR, SIS and variants. Similar to SIS, the model treated here is not microscopically irreversible, so an approach such as the one in \cite{Ale} would be impractical. Moreover, as the set of spatio-temporal configurations of contaminant sources is exponentially large, time-forward methods such as \cite{lokhov2014inferring} seem also inadapt.

While having information from multiple sensors may narrow down the search space of origins of a contamination event (by a simple intersection of the domain coverages of each sensor affected \cite{cristo2008pollution}), our approach is able to locate sources efficiently even with information coming from a single sensor.

In next section we detail the simplifications made for the forward in time contamination model. In such setting, we can write the 
inverse problem in a formal bayesian approach in section \ref{sec:bayesian}. The resulting problem is similar to standard statistical mechanics problems of spin models, and we propose a solution based on Belief Propagation algorithm in subsection \ref{sec:bp}. The bayesian approach relies on the assumption of some prior information that is parametrized with a set of ``fields''. We discuss the relation between solutions and the prior fields in a 4-node network in section \ref{sec:characterization}, passing then to testing our method on Anytown and Modena cities in \ref{sec:benchmarking} where we show that it effectively recovers the origin and the pattern even from a single sensor information.

\section{Discrete model for the contamination}\label{sec:model} 

Water distribution networks (WDN) are represented by graphs, as show in figure \ref{fig:modenaepanet},  in which nodes represent reservoirs, pumps, tanks, demand points or simply junctions of two or more pipes, while edges represent the pipes connecting them. The fluid dynamics is ruled by pressure differences along the pipes, which in turn depend on the pressure at pumps or reservoirs and the demand, corresponding within some approximations to the Todini-Pilati equations \cite{Todini}. 

\begin{figure} \centering
 \includegraphics[width=0.12\textwidth]{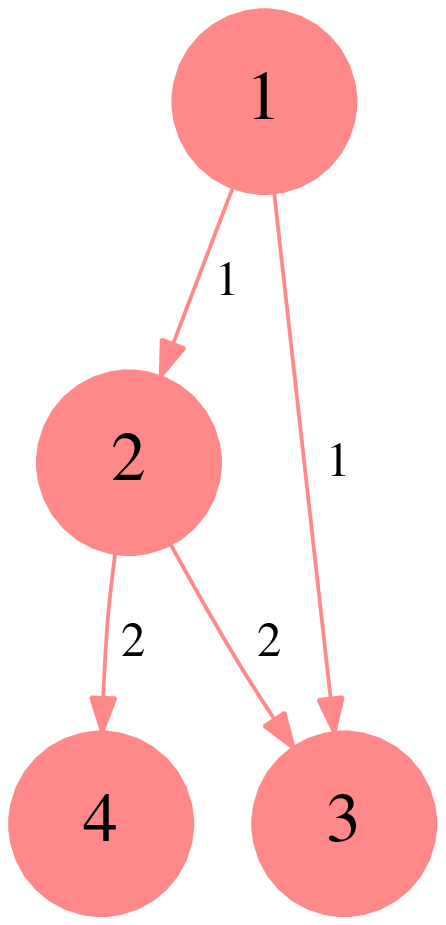}
 \includegraphics[width=0.7\textwidth]{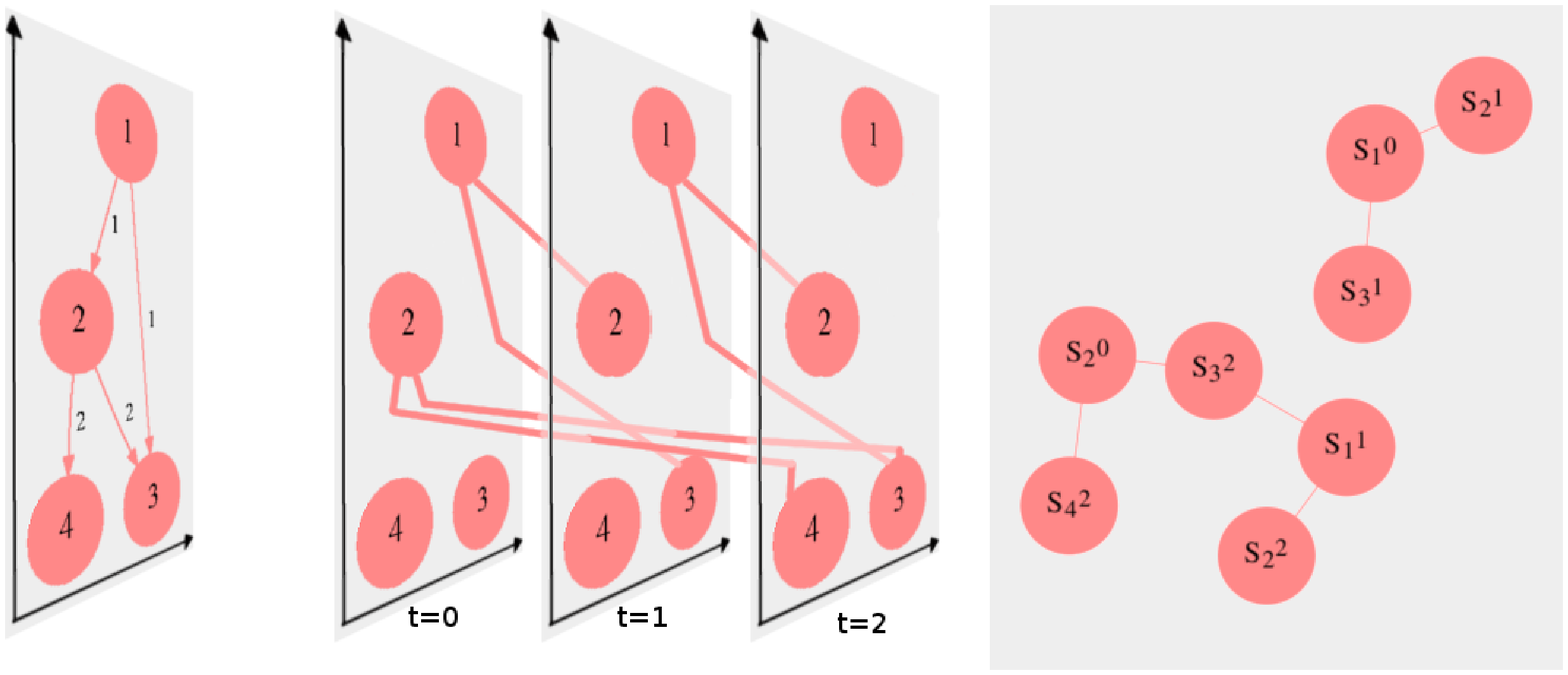}
 \caption{\label{fig:time_extended} Example of the transformation of a simple graph of a WDN into a time extended graph.  } 
% \caption{\label{extended graph}Extended graph of example network of figure \ref{fig:real graph} for a time period of analysis equal to $T$ steps. Circles are the state of a node in a time step and the connection is a direct relation between the state of two connected nodes.}
\end{figure}

While focusing on our inverse problem (contamination source detection (CSD)) we assume the direct problem of the fluid dynamics as already solved, either by simulation or by direct measuring in the real system. This means that times required for the fluid to go from one point in the network to another are known. 
If fluid velocities are known, for instance, the transport times between nodes $i$ and $j$ connected by a pipe of length $L_{i,j}$ can be computed as $\Delta_{i,j} = L_{ij}/v_{ij}$. In this sense, the water network is represented by a weighted directed graph $G=(V,E,\Delta)$ as in the first panel in Fig. \ref{fig:time_extended} where each node $i\in V$ corresponds to a physical node of the water network, and every pipe is represented by an edge $(i\to j) \in E $, with transport time $\Delta_{i,j}$.
For simplicity (but not a fundamental assumption) we stick to stationary models, those where pressures don't change in time, and therefore transport times are not changing either.

We will consider the eventual incursion of non-self-propelled contaminants into the system that are transported by the bulk of the moving fluid from the insertion point downstream through the system. Modeling of such a process in commercial software like EPANET considers non trivial aspects of its dynamics as diffusion within the fluid bulk and chemical degradation, among others. For the Bayesian modeling, however, we will keep a simplified version of this process, as done previously in \cite{paper}:
\begin{itemize}
\item discrete time: we will measure time (including pipe delays $\Delta_{i,j}$) in some discrete units;
\item binary states: any node in the network is in one of two states, clean state $0$ or contaminated state $1$;
\item non-degrading contaminant: once in the system the contaminant will propagate downstream without fading out.
\end{itemize}

In this setting, the relevant quantity for the forward dynamics of the contaminant in the network is $s_i^t\in \{0,1\}$ signaling the state (clean/contaminated) of node $i$ at discrete time $t$. This corresponds to extending our graph to consider nodes in space and time (Fig. \ref{fig:time_extended}). The edges $(i\to j)\in E$  of $G$ (the original spatial network) now are replaced by edges $((i,t)-(j,t+\Delta_{i,j})) \in E'$ in $G'$, and therefore the information about the pipe time delays is embedded in the directed graph structure. We will use the upstream neighborhood of nodes in the extended graph, defined by
\[ \deltap_{i,t} = \{ (k,t') : (k\to i) \in E , t' = t-\Delta_{k,i}\}.\]

The conservation of the contaminant implies that for every node $i$, its state in time $t$ is related to the states of the nodes connected to it $\deltap_{i,t}$ in the upstream direction $ s_i^t = 1-\prod_{\deltap_{i,t}} (1-s_j^{t-\Delta_{ji}})$. Since variables take values in $\{0,1\}$, this formula is just an algebraic mapping of the assertion that node $(i,t)$ will be contaminated if and only if at least one its upstream nodes are contaminated.
Furthermore, in our bayesian treatment of the inverse problem we will need two extra variables to describe the process of introducing contaminants in the system: 
\begin{itemize}
 \item the contaminant-pouring variable $y_i^t\in \{0,1\}$  representing the external addition of contaminants at time $t$ on node $i$, 
 \item and the contamination exposure $\upsilon_i\in\{1,0\}$ saying that node $i$ will or will not incorporate contamination coming from the exterior.
\end{itemize}
such that the forward in time behavior of the system is:
\begin{equation}
s_i^t = 1-(1-\upsilon_i  y_i^t)\prod_{\deltap_i} (1-s_j^{t-\Delta_{ji}}) %\: \equiv\: \varphi(s_i^t , \upsilon_i , y_i^t , s_{\deltap_i}^{t-\Delta})
\label{eq:state}
\end{equation}
This deterministic formula implies that the contaminant in any given point $s_i^t$ will appear for one of two reasons: either it comes from some neighboring node in the upstream network $\deltap_{i,t}$, or it is being directly poured by some external agent into the system at point $i$ and time $t$. Please notice that the exposure variable $\upsilon_i$ acts as a gate: if $\upsilon_j=0$ for some spatial node $j$, then this node will never produce contamination of its own, regardless the value of $\yit$. Also notice that there is one equation (\ref{eq:varphi}) for each node of the time extended graph $s_i^t$, and that while these variables characterize the contamination process, we are mostly interested in inferring the most likely values of $\forall_j \: \upsilon_j$ and the patterns $\left\{ \yit  : \upsilon_i =1 \right\}$, since these are the variables related to the contamination source.

\section{Bayesian approach to inverse contamination detection}\label{sec:bayesian}

Using the vector notation $\vec s_i$ to refer to all times of a given spatial node $i$, we have four sets of variables relevant to our problem:
\begin{itemize}
\item $O=\{\vec s^*_{o_1}, \vec s^*_{o_2}\ldots\vec s^*_{o_k}\}$ The sensors report on the state of the observed nodes. 
\item $S=\{\vec s_1, \vec s_2\ldots\vec s_N\}$ The state of all nodes (including those sensed).
\item $V=\{\upsilon_1, \upsilon_2\ldots\upsilon_N\}$ The variables signaling which nodes are actively introducing contamination into the system.
\item $Y=\{\vec y_1, \vec y_2\ldots\vec y_N\}$ The contamination patterns introduced at each node.
\end{itemize}
Only the values of the first set are given, and the remaining three sets are to be inferred. We can state the Source Detection Problem  problem in the following way:
\begin{defi}{}
Find the most {\it efficient} set of values for the contaminants nodes $V$ and their patterns $Y$ that are consistent with the observation of contamination in the sensor nodes $O$. 
\end{defi}
Consistent means that these values explain the observation in the sense of equation (\ref{eq:state}). Efficient means that an explanation involving less nodes $i$ as origins ($\upsilon_i=1$), and less pouring times $t$  $(y_i^t=1)$ is preferred. Efficient explanations are chosen since we fundamentally assume that contamination is a rare event, therefore an explanation with one single origin seems more credible that one involving two or more, even if both correctly reproduce the sensed contamination.

Formally, we can write the conditional probability $P(S,V, Y | O)$ of a state of the system using Bayes theorem  
\begin{equation}
P(S,V, Y | O) = \frac 1 Z P(O | S,V, Y)P(S,V, Y). \label{eq:Bayes}
\end{equation}
where $Z = P(O)$ is the (here irrelevant) normalization constant. Disregarding multiplicative constants (with respect to $S,V,Y$), the probability of the observation will be given by 
\begin{equation}
P(O | S, V, Y) \propto \exp\left(-\eta  \sum_{o,t\in O}  |s_o^t-{{s^*}_o^t}| \right)    \label{eq:PO}
\end{equation}
penalizing a discrepancy between the real states $s_o^t$ of the observed nodes and the ones reported by the sensors ${s^*}_o^t$. In the limit $\eta \to \infty$ sensors make no mistake and their report ${s^*}_o^t$ is truly the state $s_o^t$, however we will work with a big but finite $\eta > 1$.

The second factor in equation (\ref{eq:Bayes})  is (again neglecting normalization)
\begin{equation}
P(S,V, Y) \propto  \prod_{i,t} \varphi(s_i^t , \upsilon_i , y_i^t , s_{\deltap_i}^{t-\Delta}) \:\prod_{i,t} P(y_i^t) \: \prod_i P(\upsilon_i)  \label{eq:PSVY}
\end{equation} 
The first exponential term is 
\begin{equation}
 \varphi(s_i^t,\upsilon_i, y_i^t, s_{\deltap_i}^{t-\Delta} ) = \exp (-\beta  E_{i,t} )  \label{eq:varphi}
\end{equation}
with
\begin{equation}
 E_{i,t}(s_i^t,\upsilon_i, y_i^t, s_{\deltap_i}^{t-\Delta} ) = \left\{ \begin{array}{ll}
                                                                0  &  \mbox{ If } s_i^t = 1-(1-\upsilon_i  y_i^t)\prod_{\deltap_i} (1-s_j^{t-\Delta}) \\
                                                                1 & \mbox{ If inconsistent. }
                                                               \end{array} \right.  \label{eq:E}
\end{equation}
An infinite value of $\beta$ would give 0 the probability of having inconsistent processes of contamination, therefore enforcing the deterministic dynamics of eq. (\ref{eq:state}). We rather run with a high value of $\beta \sim 10^2$, which gives a very low probability (but still finite) to states that violate the deterministic dynamics. 

The last two factors in (\ref{eq:PSVY}) are the priors $P(\upsilon_i)$ and $P(\yit)$ parameterized with the fields $\lambda$ and $\gamma$:
\begin{eqnarray}
P(\upsilon_i) &\propto &\exp^{-\lambda \upsilon_i} \label{eq:P_u}\\
P(\yit) &\propto &\exp^{ -\gamma \yit} \label{eq:P_y}.
\end{eqnarray}
Field $\lambda$ defines the prior probability for having node $i$ actively introducing contaminant in the network, while $\gamma$ gives the probability of finding contamination being introduced at a particular time $t$. The fact that both fields are positive is the mathematical expression of our basic assumption, i.e. that contamination is a rare event. Finally, our posterior probability is given by:
\begin{equation}
P(S,V,Y | O) = \frac 1 Z e^{(-\beta\sum_{i,t}E_{i,t} +\sum_{i}\lambda \upsilon_i +\sum_{i,t}\gamma \yit +\sum_{o,t}\eta|s_o^t-{s^*}_o^t|)}.  \label{eq:Ptotal}
\end{equation}
which has a familiar Boltzmann probability distribution form. In statistical mechanics jargon, the observation acts as a fixed (quenched) disorder and the dynamical variables are $S,V,Y$. This probability measure (and all the relevant moments) depend on the parameters we have introduced:
\begin{itemize}
\item $\beta$ A temperature like parameter enforcing the consistency of the contamination dynamics in the network. 
\item $\eta$ A field penalizing discrepancies between the observed state and the real state of the observed nodes.
\item $\lambda$ A field forcing the contamination to be a rare event (in space).
\item $\gamma$ A field forcing the contamination to be a short event (in time).
\end{itemize}

To the effect of the contamination source detection, we are mostly interested in $\ui$, which will take value $\ui = 1$ for nodes $i$  that are the source of contamination. We, therefore, want to compute $\langle \ui \rangle$ over the measure defined by eq. (\ref{eq:Ptotal}), which we intend to compute using belief propagation \cite{yedidia2005constructing}, and we will infer as the origins of the contamination the nodes $i$ such that $\langle \ui \rangle > 0.5$.

\subsection{Belief Propagation}\label{sec:bp}

In the exponent of the measure (\ref{eq:Ptotal}) complication arises from the energetic part $\sum_{i,t}E_{i,t}$, since the energy defined in (\ref{eq:E}) is the only element of this measure that is not fully factorized over variables. In absence of this term, of course, every variable would be independent and $P(S,V,Y|O) = \prod_{i,t} b_{i,t}(\sit) \prod_{i,t} b'_{i,t}(\yit) \prod_i b_i(\upsilon_i) $ and moment computation would be trivial. However, the interactions given by the dynamics of the contaminants in the network makes the measure (\ref{eq:Ptotal}) a non factorized one. Computing marginals naively can be an impossible task, since the direct marginalization has a cost $O(2^N)$, where $N$ is the number of binary variables in the model. We will resort to an approximative variational method called Belief Propagation \cite{yedidia2005constructing}. 

\begin{figure} \centering
  \includegraphics[width=0.35\textwidth]{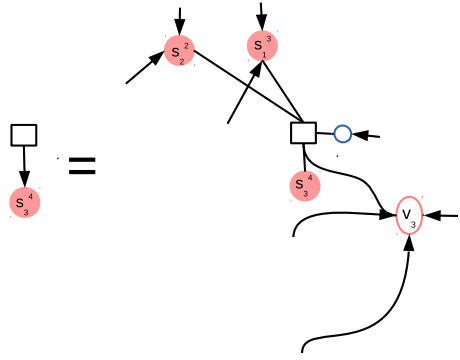}
\caption{\label{fig:factor_graph_message} Graphical representation for the update equation \ref{eq:ba}.}
\end{figure}

We can not reduce $P(S,V,Y|O)$ to a set of factorized probability measures $\{ \forall_{\sit \in V'} b_{i,t}(\sit) \}$ ,$\{\forall_{ \yit \in V'} b'_{i,t}(\yit)\} $,  $\{\forall_{\upsilon_i \in V'} \prod_i b_i(\upsilon_i) \}$. But we can try to represent the full distribution by all this factorized measures plus a collective measure for the small groups of variables that interact in every energetic term (\ref{eq:E}). Lets simplify the notation a bit by relabeling $x_q$ with $q \in [1\ldots N]$ all the variables of our problem, indistinctly of whether they are s-type, y-type or $\upsilon$-type. Let us also relabel $a_q$, in agreement with the factor graph representation of the appendix 1, every energetic term $E_{i,t}$, and $\partial a_q \equiv x_a = \{s_i^t,\upsilon_i, y_i^t, s_{\deltap_i}^{t-\Delta} \}$ the set of all the variables entering that interaction (e.g. Fig. \ref{fig:factor_graph_message}). We can attempt a description of the model by considering also the distributions over every set of interacting variables $\{ \forall_a b_a(x_a) \}$. 
%, the idea is to consider a large set of measures  where the measure itself is fractioned into one-variable-measures $b_i(x_i)$ and interaction-measures $b_{a} = (x_1,x_2,\ldots)$ that are consistently computed.
%Using the factor graph representation explained in the appendix section, allow for a direct implementation of the message passing method known as belief propagation . 

Belief Propagation \cite{yedidia2005constructing} consists in a system of fixed-point equations for positive real quantities (messages) $m_{a \to i}(x_i)$ (see appendix 2 for more details). From a fixed point, approximated marginal distribution ``beliefs'' can be computed  as
\begin{eqnarray}
 b(s_i^t) & \propto& \displaystyle \prod_{a \in \partial {\sit}} m_{a \to i}(\sit) \nonumber \\
 b(y_i^t)& \propto&  e^{-\gamma \yit} \displaystyle m_{a[i,t] \to y}(\yit)  \label{eq:bi} \\
 b(\ui)& \propto&  e^{-\lambda \ui} \displaystyle \prod_{a \in \partial {\ui}} m_{a \to i}(\ui)  \nonumber
\end{eqnarray}
while the belief over every set of interacting variables is given by
\begin{equation}
b_a(\sit,\yit,\upsilon_i,s_{\deltap_i}^{t-\Delta}) \propto \exp{\left(-\beta E_{i,t} - \lambda \upsilon_i - \gamma \yit - [ \eta|s_i^t-{s^*}_i^t| ]  \right)} \displaystyle \prod_{x=\sit,\upsilon_i,s_{\deltap_i}^{t-\Delta}}\displaystyle \prod_{c \in  \partial x \setminus a} m_{c\to x}(x)   .  \label{eq:ba}
\end{equation}
where the term between square brackets is only present if the node is among the observed (sensed) nodes $O$.
The equations for the messages are obtained by iteratively trying to satisfy the consistency that they are supposed to enforce (\ref{eq:marginalization}). For instance, by equating the first equation in (\ref{eq:bi}) with (\ref{eq:ba}), we obtain
\begin{eqnarray}
m_{a\to i}(s_i^t) &\propto \displaystyle \sum_{\yit,\upsilon_i,s_{\deltap_i}^{t-\Delta}\setminus i} \exp{\left (-\beta E_{i,t}- \lambda \upsilon_j -\gamma \yit \right)}\displaystyle \prod_{x= \upsilon_i,s_{\deltap_i}^{t-\Delta}\setminus s}\displaystyle \prod_{c \in  \partial x \setminus a} m_{c\to x}(x)  \label{eq:mensaje}
\end{eqnarray}
A graphical representation for this update equation, for instance, would look like Fig. \ref{fig:factor_graph_message} (using the same example of the appendix). The equations for $m_{a_{i,t} \to \upsilon_i}$, $m_{a_{i,t} \to  y_i}$ and $m_{a_{i,t} \to s_j^{t-\Delta_{j,i}} }$ are similarly derived.

The overall approach then consists of the following steps (see also algorithm \ref{alg:bp} in appendix 2):
\begin{enumerate}
 \item Construct a time extended graph for the Water Distribution Network, as schematized in Fig. \ref{fig:time_extended}.
 \item Construct a factor graph representation for the Bayesian inverse problem (as explained in the appendix).
 \item Iteratively update the self-consistent equations for the messages in the network (similar to (\ref{eq:mensaje})), until they stop changing.
 \item When (if) converged, evaluate the probabilities eq. (\ref{eq:bi}).
 \item Propose as origing of the contamination nodes with high probability $b_i(\upsilon_i = 1)$.
\end{enumerate}

%It is costumary to interpret the Lagrange multipliers as messages from factor nodes to variables telling them the probability of each state, according to the interaction and the messages received by the rest of the variables. 

%The first step in the implementation is to read the information of an extended graph from a file, so we have now the information of which variable nodes of the real graph are present in our problem and which other variable nodes are connected with its. Then the variable nodes of $\yit$ and $\ui$ variables are created in that order with its connections and are added to the list of variable nodes, at the end factor nodes are created for each variable node from the real graph with a variable for save the message it going to send to its neighbors. 

% \begin{eqnarray*}
%  b(x_i) \propto \displaystyle \prod_{a \in \partial_{i}} m_{a \to i}(x_i)  = \exp( \sum_{a \in \partial_{i}} u_{a\to i}) = \exp(H_i)
%  b(s_i^t)& \propto \displaystyle \prod_{a \in \partial {\sit}} m_{a \to i}(\sit) &= \exp( \sum_{a \in \partial {i}} u_{a\to i} \sit) = \exp(H_{i,t} \sit)\\
%  b(y_i^t)& \propto  e^{\gamma \yit} \displaystyle m_{a[i,t] \to y}(\yit) &= \exp( \gamma \sit + u_{a[i,t]\to i} \sit) = \exp(H_{-i,t} \yit)  \\
%  b(\ui)& \propto   e^{\lambda \ui} \displaystyle \prod_{a \in \partial {\ui}} m_{a \to i}(\ui)&= \exp( \lambda\ui  +\sum_{a \in \partial {i}} u_{a\to i} \ui) = \exp(H_i \ui) 
% \end{eqnarray*}

\section{Characterization in simple examples} \label{sec:characterization}

\begin{figure} \centering
\begin{center}
 \includegraphics[width=0.15\textwidth]{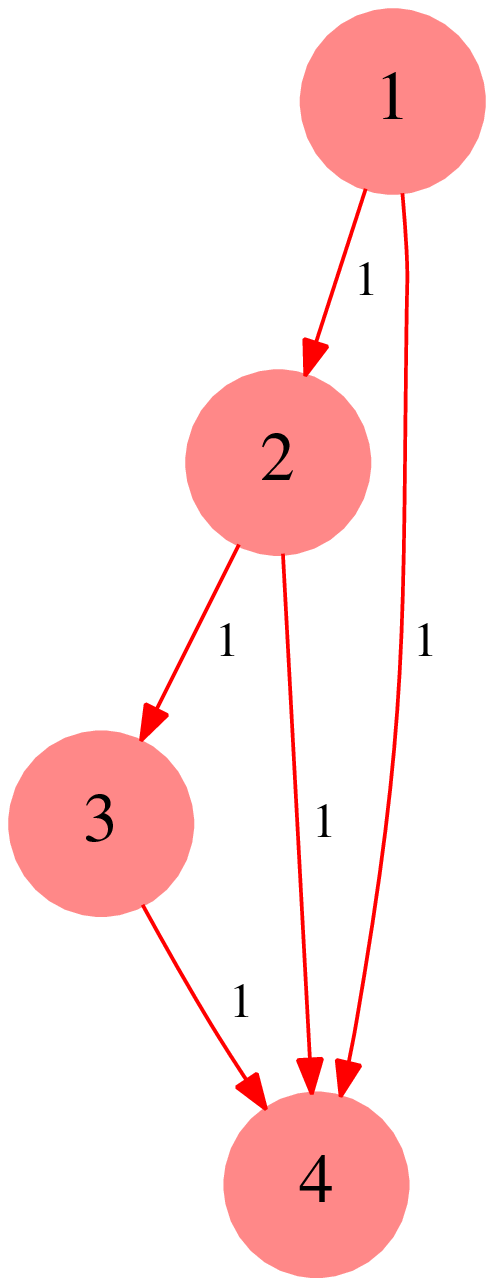}
 \includegraphics[width=0.36\textwidth]{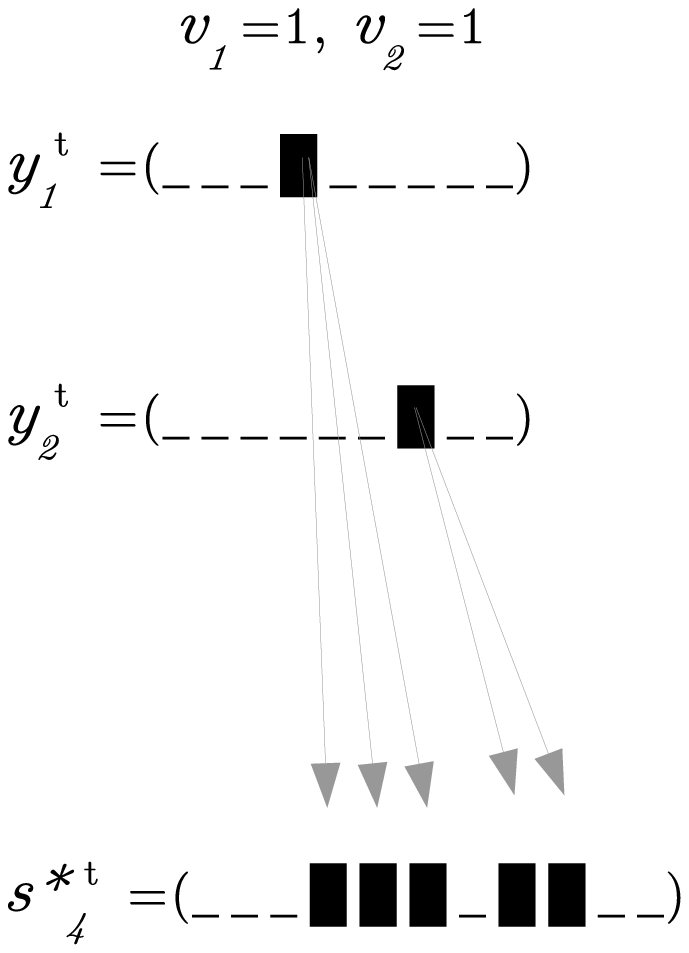}
 \includegraphics[width=0.36\textwidth]{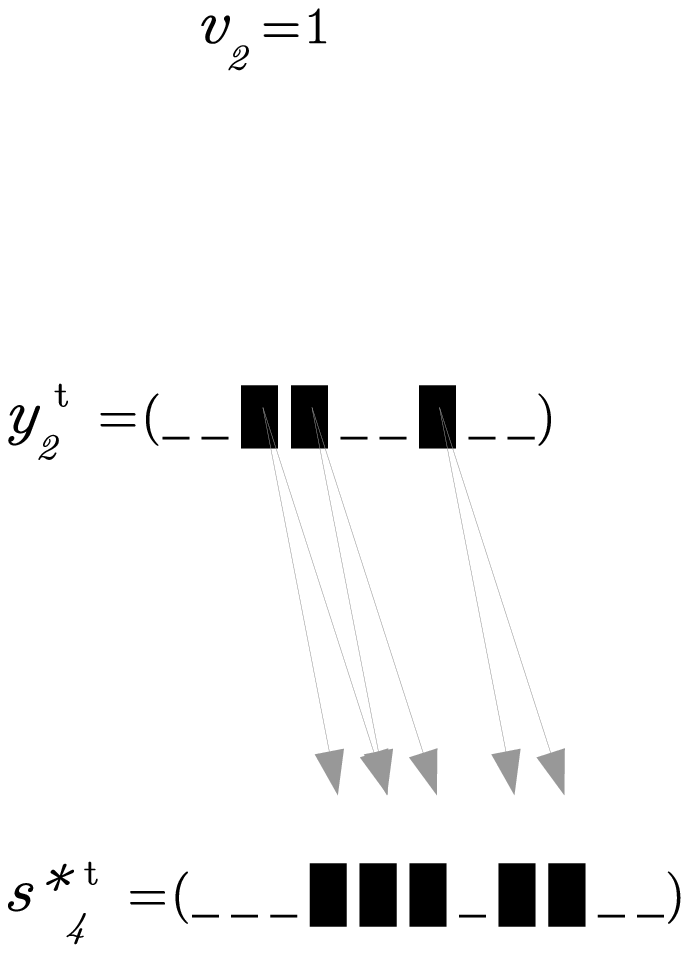}
\caption{\label{fig:3caminos} Representations of optimal spills explaining the observation. Spill at two nodes with 1 time contaminated at each. Spill in only one node with t}
 %\caption{\label{extended graph} Factor graph of example network of figure \ref{fig:real graph} for a time period of analysis equal to $T$ steps. Circles are the variable nodes and box are factor nodes, the connections are only between interaction and variable nodes.}
\end{center}
\end{figure}

Example in figure \ref{fig:3caminos} is handy to understand the whole procedure and the role of the new parameters  $\lambda$, $\gamma$ and $\eta$ in the solution of CSD problem. Let us assume that a pattern ${s^*}_4^t = (0,0,0,1,1,1,0,1,1,0)$ is detected at sink node $4$, at the bottom of the network. Simple although not optimal explanations for the sensors report is that the node $4$ itself, or the node $3$ are the origin of the contamination with a pattern $y_i^t$ similar to the one detected. In either case we require 5 times for the active contamination event.

There are two explanations more parsimonious as shown in figure \ref{fig:3caminos}. One involving two nodes (1 and 2) and two moments for the contaminant to be pushed into the system, and one other involving only node (2) with 3 moments in which contamination is being poured. This two explanations are more efficient than those with nodes $4$ or $3$. Which of them our algorithm would prefer, however, depends on the relative weight given to the fields $\lambda-\gamma$ that control space-time parsimony. 

\begin{figure} \centering
 \subfigure[Nodo 1]{\includegraphics[width=0.4\textwidth]{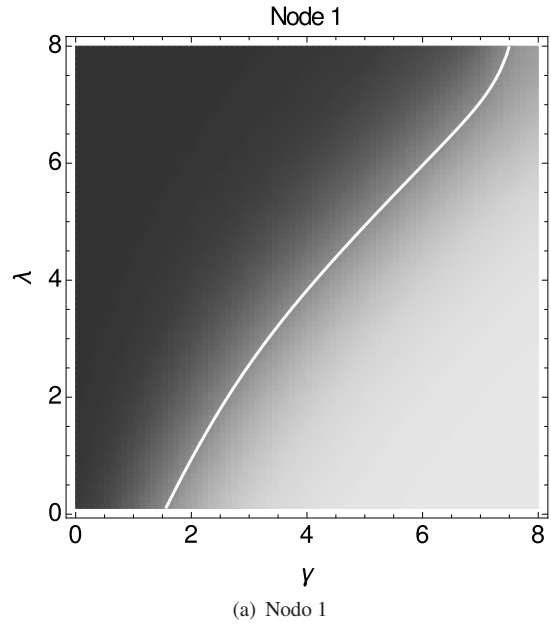}}
% \subfigure[Nodo 4]{\includegraphics[width=0.25\textwidth,angle=270]{2nodos_4_gamm_1.eps}}
 \caption{\label{fig:belief_node_1} Probability $b_{1}(\upsilon_{12})$ for nodes $1$ to be the origin of the contamination observed at node 4 in Fig. \ref{fig:3caminos}, for different values of parameters $\lambda$ and $\gamma$ with fixed $\eta = 8.0$. White isolines mark the $b(\upsilon) = 0.5$ boundary, and dark colors are the lower probabilities.}
 %\caption{\label{extended graph} Factor graph of example network of figure \ref{fig:real graph} for a time period of analysis equal to $T$ steps. Circles are the variable nodes and box are factor nodes, the connections are only between interaction and variable nodes.}
\end{figure}

After convergence, our algorithm forecast low probabilities for nodes $3$ and $4$ of being the origin of the contamination, and high probability $b_2(\upsilon_2=1) \simeq 1$ for node $2$, for all fields values tested. As of node $1$, figure \ref{fig:belief_node_1} shows the probability $b_1(\upsilon_1)$ to be an active contamination node for different values of $\lambda$ and $\gamma$. Roughly speaking, whether $\lambda>\gamma$ or not defines the type of solution found in this case, from one that is optimal in space (only node 2 is origin) to one that is optimal in time (nodes 1 and 2 are origin, but only with 2 moments of contamination).

\begin{figure} \centering
\subfigure{ \includegraphics[width=0.3\textwidth,angle=0]{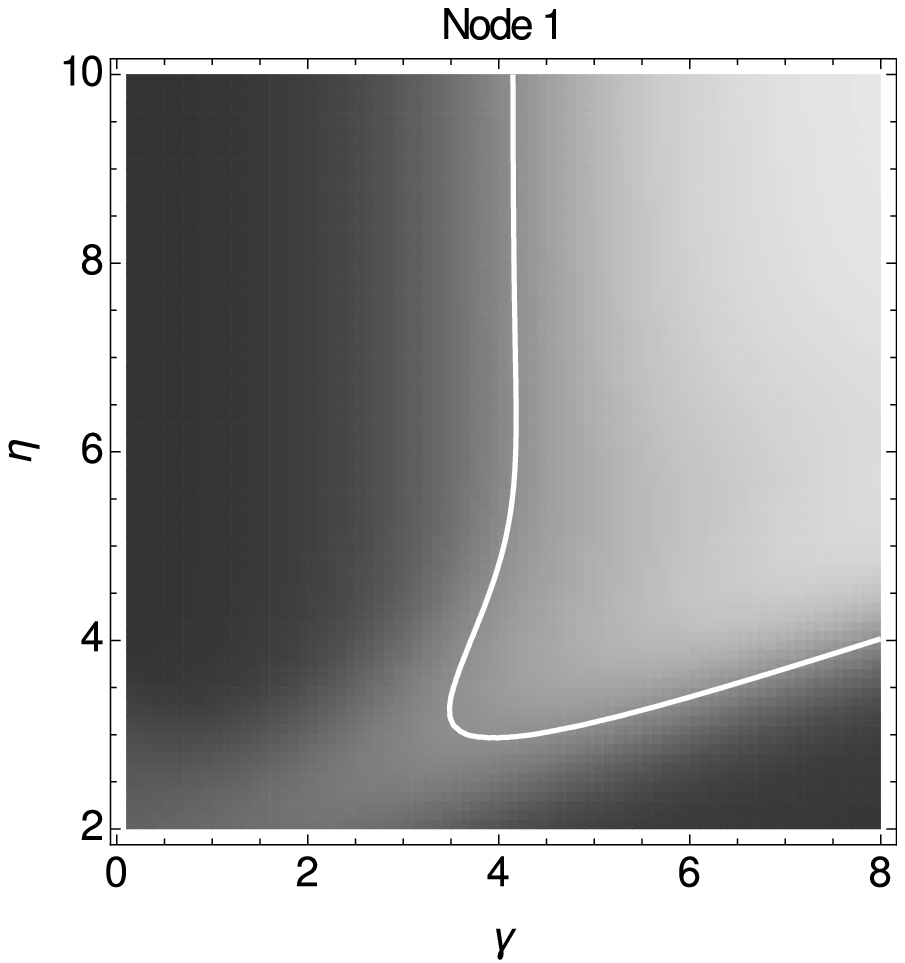}} \hspace{1cm}
\subfigure{ \includegraphics[width=0.3\textwidth,angle=0]{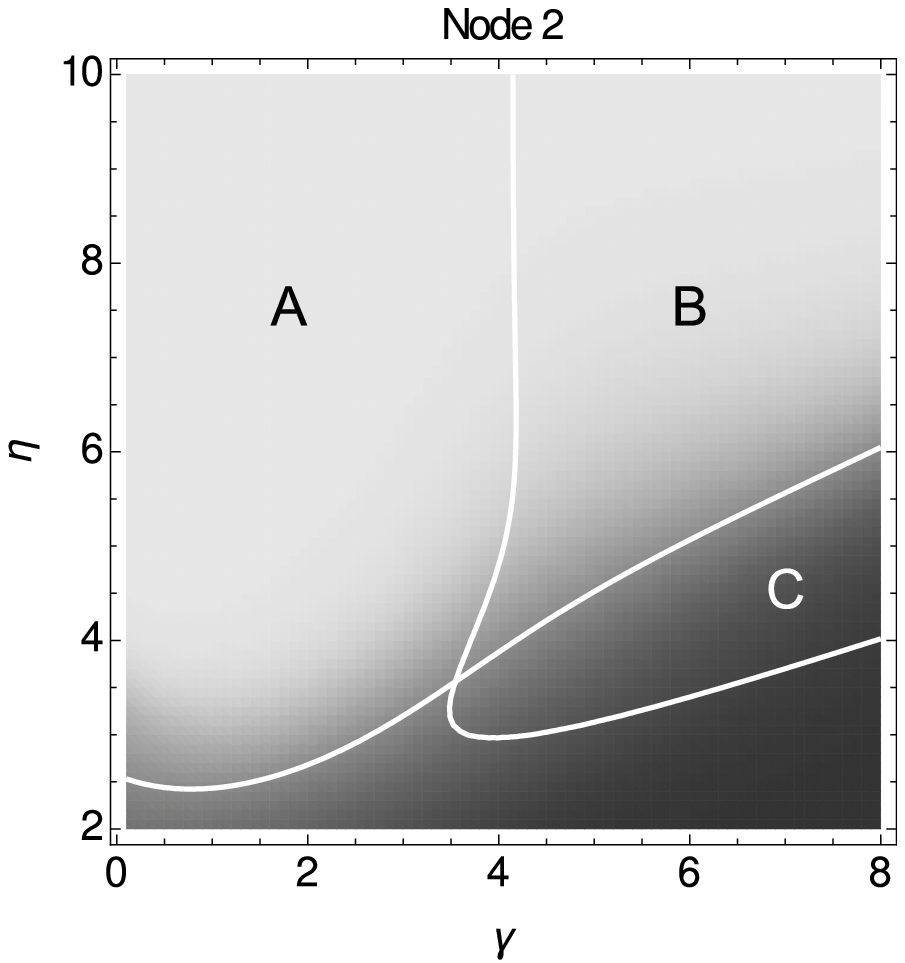}} % \hspace{0.35cm}
%\subfigure{ \includegraphics[width=0.035\textwidth,angle=0]{colorscale.pdf}}
 \caption{\label{fig:3caminos_eta} Probabilities $b_{1,2}(\upsilon_{1,2})$ for each of the nodes $1$ and $2$ to be the origin of the contamination observed at node 4 in Fig. \ref{fig:3caminos}, for different values of parameters $\eta$ and $\gamma$ with fixed $\lambda = 4.0$. White isolines mark the $b(\upsilon) = 0.5$ boundary, and dark colors are the lower probabilities. In Node 2 panel we also show the Node 1 isoline. Different solutions A,B and C are explained in the main text.}
\end{figure}

A similar interplay is seen between the $\gamma$ and $\eta$ parameters, at fixed $\lambda = 4.0$, in figure \ref{fig:3caminos_eta}. Let us remember that $\eta$-field enforces the dynamics to fully reproduce the sensor's report (see eq. (\ref{eq:ba})), and a low value of $\eta$ allows the inference procedure to maximize space or time in the contamination source by sacrificing the consistency with the observation. The second panel in Fig.  \ref{fig:3caminos_eta} is divided in regions:
\begin{enumerate}
 \item[\bf A] corresponds to the solution with only node $2$ as origin, since the field $\gamma$ that forces time-optimal solutions, is low.  
\item[\bf B] corresponds to both nodes 1 and 2 being the origin since $\gamma$ is bigger. 
\item[\bf C] is interesting: a solution with only node $1$ as the origin, at the expense of not reproducing the last two signals observed at the sensor in node 4. See also the central panel in figure \ref{fig:3caminos}
\end{enumerate}
Fig.  \ref{fig:3caminos_eta} also shows that outside of regions A, B and C, corresponding to low values of fidability $\eta<3$ that produce a trivial answer: since contamination is such a rare event and you can not trust the sensor too much, just consider there has been no contamination whatsoever!

When using belief propagation for the contamination source detection problem we should bear in mind the role of these fields $\lambda,\gamma$ and $\eta$, since as this simple examples show, by varying them we can select different solutions. Instead of considering this flexibility a handicap of the procedure, we retain it as a relevant feature. As will be seen in the last section, the fact that we can disregard some of the information given by the sensor, allows the method to be useful even when tested again realistic simulations of contamination that do not exactly fit the that assumed for the forward model.

%\begin{figure} \centering \begin{center}
% \includegraphics[width=0.4\textwidth,angle=270]{2nodos_prob.eps}
% \caption{\label{fig:3caminos_eta_both} Belief on of nodes 2 and 4 from graph \ref{fig:3caminos} versus parameter $\eta$ for $\lambda = -8$ and $\gamma = -4$.}
 %\caption{\label{extended graph} Factor graph of example network of figure \ref{fig:real graph} for a time period of analysis equal to $T$ steps. Circles are the variable nodes and box are factor nodes, the connections are only between interaction and variable nodes.}
 %\end{center}\end{figure}
 
%Note that as it was point belief on of node 2 is higher than belief on of node 4 in the middle region corresponding to $4 < \eta > 6$ it is an interesting fact is necessary analyze. Lets focus again in the detected contamination pattern $0,0,1,1,1,0,1,1,0,0$ as was remarked are two moments of contamination one can be explain optimally by a spill in node 2 and the other can be explain optimally by a spill in node 4, but while $\eta$ is decreasing is more likely a mistake in sensor measurement, so it is logical consider one of sensor detection is wrong. If is considered detections in times 6 or 9 was wrong then second part of pattern can be explained optimally by node 2 so the all pattern can be explained by node 2 so it will be the mos probable source of the contamination.         

\section{Benchmarking on Anytown and Modena} \label{sec:benchmarking}

In this section we will present some examples of the application of the inference method in two different networks Anytown and Modena.

\subsection{Anytown}

\begin{figure} \centering
 \includegraphics[width=0.4\textwidth]{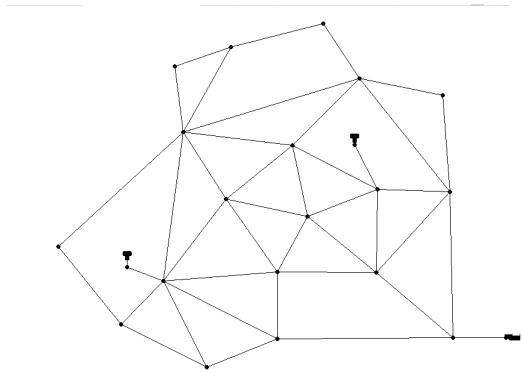}
\caption{\label{fig:anytown}  Anytown toy network visualized with EPANET.}
\end{figure}

\begin{figure} \centering
\subfigure{ \includegraphics[width=0.25\textwidth]{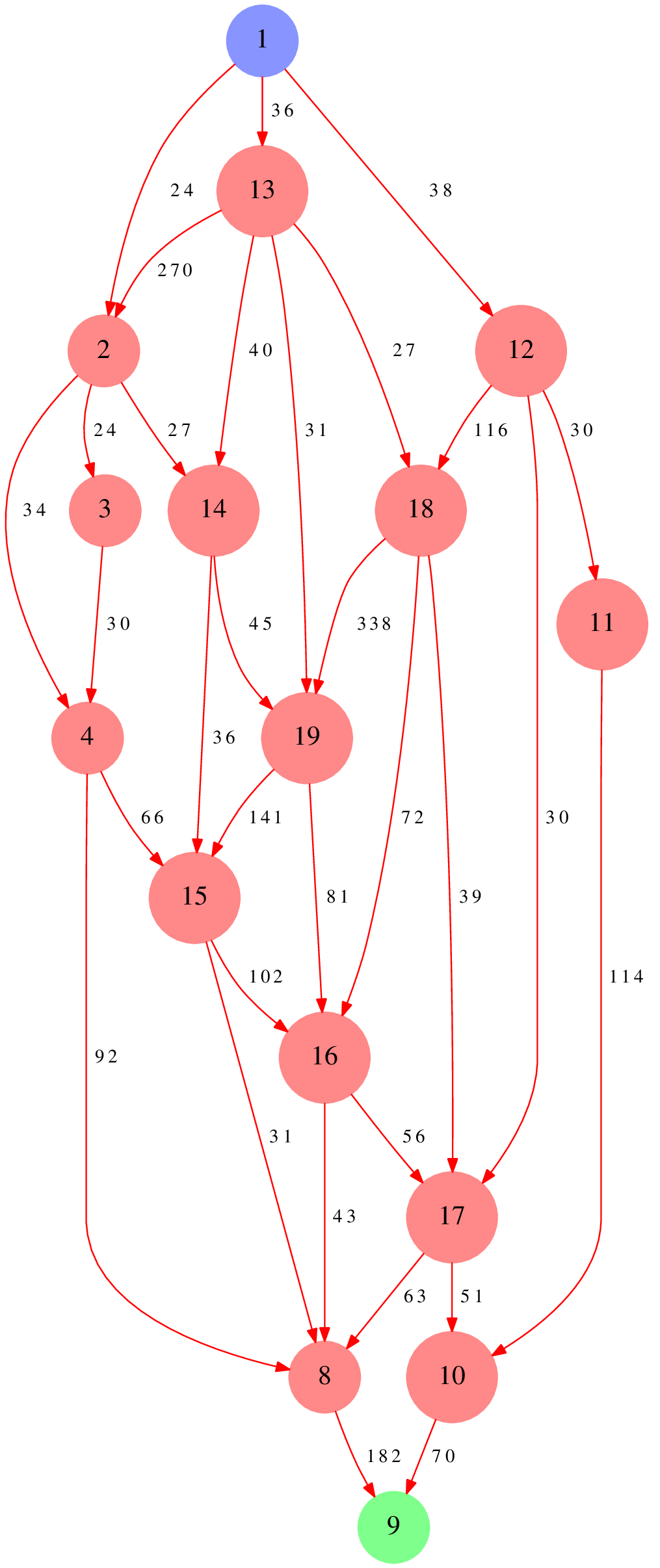}} \hspace{1cm}
 \subfigure{ \includegraphics[width=0.42\textwidth,angle=0]{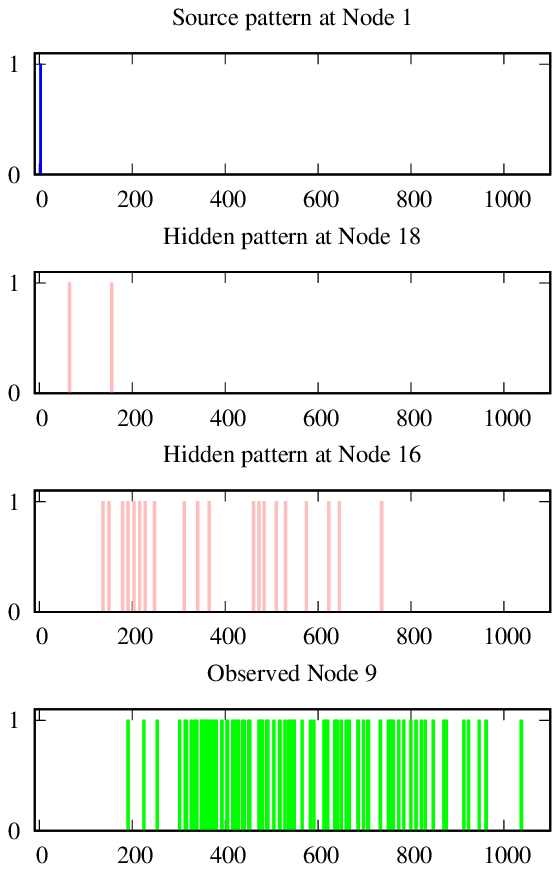} } 
\caption{\label{fig:spill} {\bf Left panel:} directed weighted graph representation of Anytown network Fig. \ref{fig:anytown}. Blue node 1 is the source of contamination, and a sensor has been placed in the green node 9. {\bf Right panel:} temporal patterns for the contamination as it moves from the source to the sensor node. Reconstruction using belief propagation is perfect in this case, recovering the correct source and the correct patterns. }
\end{figure}

In the figure \ref{fig:anytown} is shown the network of the WDN of Anytown viewed with EPANET. Being Anytown a small example, we tried our inference procedure by sensing the most downstream node 9 a contamination that entered the system at the most upstream node 1. In figure \ref{fig:spill} we show the directed weighted graph of the part of the Anytown network that could originate a contamination in node 9 (notice nodes 5,6 and 7 are not present). This is the part of the nodes that belong to the domain of coverage of node 9, using terms from \cite{cristo2008pollution}. 

Sensing and origin nodes are colored green and blue respectively. The times $\Delta_t$ (in minutes) that contaminants take to propagate along each pipe are obtained from the EPANET output of a stationary regime in Anytown WDN with standard parameters of pressure and demand. A simple proxy for such times is $\Delta_{ij} = L_{ij}/v_{ij}$, where $L_{ij}$ is the length of the pipe between nodes $i$ and $j$, and $v_{ij}$ is the mean speed of the fluid as obtained from the solution of the Todini-Pilati equations, for instance. However, we took the values of $\Delta_t$ ``experimentally'' by considering contaminants in each node of the network, and measuring the time in which they appear in their neighbors nodes.

Panel in the right of figure \ref{fig:spill} shows the source pattern, which is consistent with the basic assumptions made in this work: is poured in few places (actually only node 1) and it is of short duration (5 time bins). We used a discretization of time in minutes for this network. The corresponding factor graph contained $25 588$ nodes of type $s_i^t$ and the same amount of $y_i^t$ nodes. We needed $n=16$ $\upsilon_i$ nodes, one for each node involved in the contamination of node $9$.

%\begin{figure} \centering
% \includegraphics[width=0.08\textwidth,angle=270]{real_anytown_19.eps}
%\caption{\label{fig:anytown_19}  Pattern of contamination detected as node 19.}
%\end{figure}

\subsubsection{Simplified dynamics}

We will first propagate this pattern down the network with the simplistic dynamics of equation (\ref{eq:state}). The resulting observation made at sensor node 9 is plotted in green at the bottom of the right pannel. We run our inference algorithm with standard parameters $\eta = 100.0, \lambda=4, \gamma=4$, and it converges with $\mbox{diff} f= \max(|m_{old} - m_{new}|)<10^{-7}$ after $37$ iterations in $20$ seconds on a desktop (4x Intel Core i5, 3.2GHz). As expected, the algorithm recovers correctly the origin node with $b_1(\upsilon_1 = 1) = 1$, and the remaining $\forall_{k\neq 1} b_k(\upsilon_k)<10^{-7}$ for all other nodes. It also recovers the contamination pattern at the source with $b_1(y_1^t )\sim 1$ only for the values of $0<t<6$. The contamination observed at other nodes in the system is also faithfully reproduced from $b_i(s_i^t)$. We don't plot any of these results in the right panel of Fig. \ref{fig:spill} since in this scale they are not distinguishable from the true patterns.

\subsubsection{Realistic EPANET simulation in Anytown}
Real contamination obeys less neat laws than the simple dynamics of eq. (\ref{eq:state}). For instance, contaminants diffuse inside the fluid, and therefore a deltaic in time event of contamination spreads over more than one time interval when it propagates through the pipes. Furthermore, for this very reason, if not others, its concentration decays, and eventually it might go below sensors sensitivity. Therefore it is expected that our inference method will have a hard time trying to reconstruct the origin of contamination when paths are long since the reported sensed pattern might be inconsistent with the simplified dynamics assumed.

\begin{figure} \centering
\includegraphics[width=0.45\textwidth]{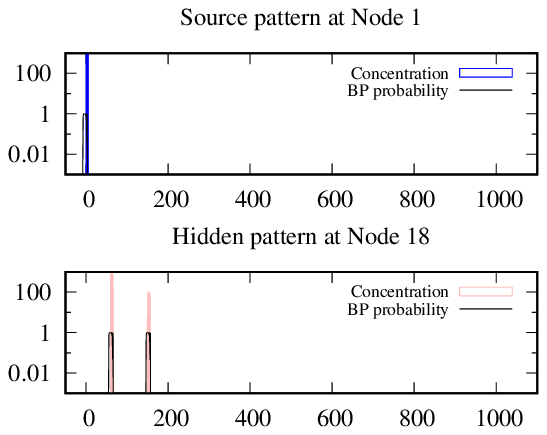}
\includegraphics[width=0.45\textwidth]{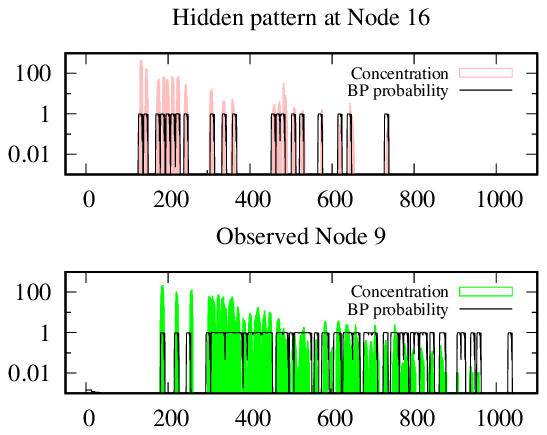}
\caption{\label{fig:epanet_anytown} Comparison between temporal patterns for the contamination as it moves from the source to the sensor node. Reconstruction of contamination pattern using belief propagation vs real patterns of contamination obtained from the direct process simulated with EPANET.}
\end{figure}

To test the resilience of our method, we start from the same source contamination of the previous example (5 times bins in node 1) but we simulate its propagation with EPANET. The real contamination $C_9(t)$ at sensor 9 is shown in green in figure \ref{fig:epanet_anytown}. Since our model assumes a binary state for the sensor, we consider the following definition for the sensor state:
\[ {s^*}_9^t = \left\{ \begin{array}{ll}
                        1 & \mbox{if EPANET reports } C_9(t)>0 \\
                        0 & \mbox{if } C_9(t) =0.
                       \end{array}
\right.\]
The realistic simulation takes into account diffusion as well as chemical decay of contamination (the nature of the contamination is irrelevant for us). Therefore, the largest path (say $T>1000$) connecting node 1 with node 9 do not contribute with a measurable amount of contaminant at node 9 and EPANET reports 0 contamination in times where there should have been.

Fortunately the Bayesian approach used allow us to give lower relevance to the sensed reports compared to that given to short explanations. This means that we can run our algorithm with typically low value for $\eta = 2$ while having large fields $\lambda = 6$ and $\gamma = 6$. The factor graph in this case was made of $30 852$ nodes $ s_i^t$ and the same amount for the $y_i^t$ and only $16$ $\upsilon_i$ nodes. The algorithm converged in $228$ iterations after $2$ minutes $10$ seconds and reports correctly node 1 to be the origin with high expectation $b_1(\upsilon_1=1)=1$ while giving $\forall_{k\neq 1} b_k(\upsilon_k)<10^{-3}$ for the other possible origins. Consistently with the simplifications of our model, the predicted pattern at the observed node obtained from $b_9(s_9^t)$ differs with the real report ${s^*}_9^t$ of the sensors (Fig. \ref{fig:epanet_anytown}), showing that the algorithm indeed chose an efficient explanation at the cost of considering sensors not fully reliable. Plots for nodes $18$ and $16$ are examples of the matching between the real concentration in those nodes, and the probability of being contaminated $b_{16,18}(s_{16,18}=1)$ as obtained from our algorithm. 

The solutions found depend on the set of parameters chosen. Running the algorithm with $\eta = 8$, for instance, will report two nodes as origins,  instead of the true unique source node 1. This example also shows a typical drawback of message passing and belief propagation equations: if we run with some other set of parameters (say $\lambda = 4, \gamma = 4$) the algorithm gets confused by the impossibility to satisfy all the equations and does not converge.

\subsection{Modena}

\begin{figure}
\centering
\includegraphics[width=0.18\textwidth]{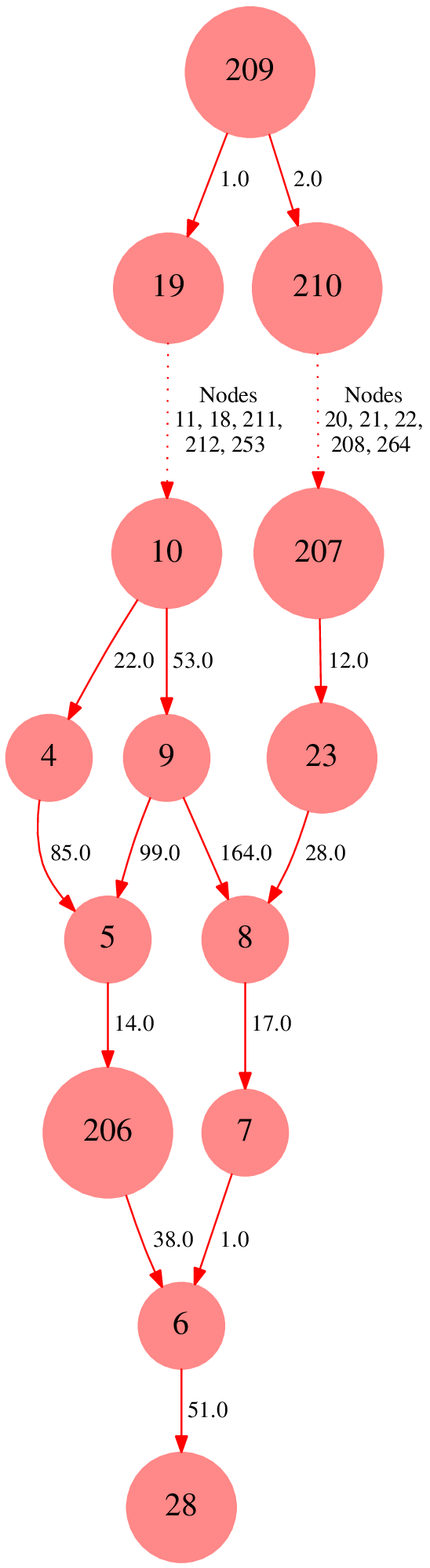}
\includegraphics[width=0.55\textwidth]{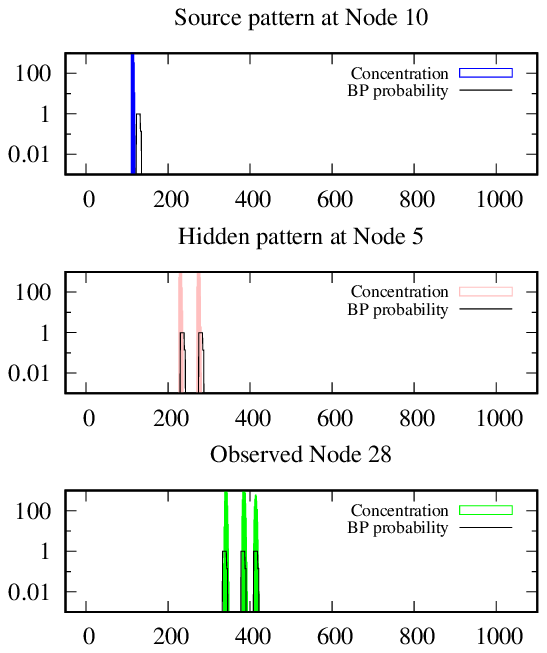}
\caption{\label{fig:modena_inference} {\bf Left panel:} directed weighted graph representation of shadow part of Modena city. \ref{fig:modenaepanet}. {\bf Right panel:} comparison between temporal patterns for the contamination as it moves from the source to the sensor node. Reconstruction of contamination pattern using belief propagation vs real patterns of contamination obtained from the direct process simulated with EPANET. }
\end{figure}

Finally we try to infer a deltaic contamination pattern on a real city network, that of Modena, in Italy, using the network and flow parameters provided in \cite{Modena}. In Fig. \ref{fig:modenaepanet} the position of a sensor at node 28th is marked in green, while the area of the network that connects with this location is marked in gray. We located a deltaic in time contamination at node 19th, which is just after the elevated tank. We simulated the propagation with EPANET, and detected the signal shown in the right-bottom plot of figure Fig. \ref{fig:modena_inference} at the sensor. With this information we try to individualize the source of contamination from all candidate nodes  (left panel in Fig. \ref{fig:modena_inference}).

Our inference method run with parameters $\gamma=\lambda=6$ and $\eta=4$ signaling a high probability for node $10$ to be the only origin of a deltaic contamination event. We compare the actual concentration at node 10 and node 5 and the (binary) one predicted by our method. In spite of the node 10 not being the correct origin, the method does predict the correct patterns in time at the intermediary nodes of the process. The reason why the algorithm produces 10 and not 19 as the origin is clear: they are connected by a single line of nodes, and therefore there can not be any efficiency distinction between 10 and any of the nodes up this chain up to nodes 19 and 209. In general, every time a chain like this appears, it has to be understood that a high probability for any of this nodes to be the origin is also applicable to any node in the chain.

\section{Conclusions}

We have developed a Bayesian approach to infer the origin of a contamination event from the information gathered in few sensor nodes in a water distribution network. The method relies on some strong simplifications, the stronger of which is considering only binary (0/1 = clean/contaminated) states in the nodes of the network. It completely disregards all the relevant information coming in the detected real-valued concentration. It also relies on a fundamental assumption: contamination events are short events coming from few nodes.

Despite this simplifications, the method is quite efficient in recovering the source of contamination in the tested toy networks. Furthermore, its probabilistic character allows for certain flexibility that achieve correct results even for realistic networks with realistic contaminant diffusion dynamics.

This method substantially improves over our previous work \cite{paper} where the solution coming from  integer linear programing used to be highly degenerate. The bayesian inference we implemented results in a single probabilistic description (that could represent several individual solutions). Furthermore, the requirement that the solution to involves as few nodes as possible is directly embedded into the prior, concentrating the posterior probability on such configurations. We pay the prize of having at least 3 parameters to handle and facing convergence problems when the model is set to reconstruct complicated patterns (in the sense that they don't follow the simple dynamics assumed).

We consider that this work opens the approach of statistical mechanics to CSD. Along this path, the next step would be to treat the concentration information in the same bayesian setting by relaxing the binary variables to take continuous values.

\section{Acknowledgements}
Work supported by the European Union’s Horizon 2020 research and innovation programme MSCA-RISE-2016 under grant agreement No 734439 INFERNET.
x
\pagebreak

\appendix
%{Appendix: Factor graph representation}

\subsection{Appendix 1: Factor graph representation}
\label{ap:factorgraph}

\begin{figure}
\centering
 \includegraphics[width=0.13\textwidth]{small.eps}
   \includegraphics[width=0.4\textwidth]{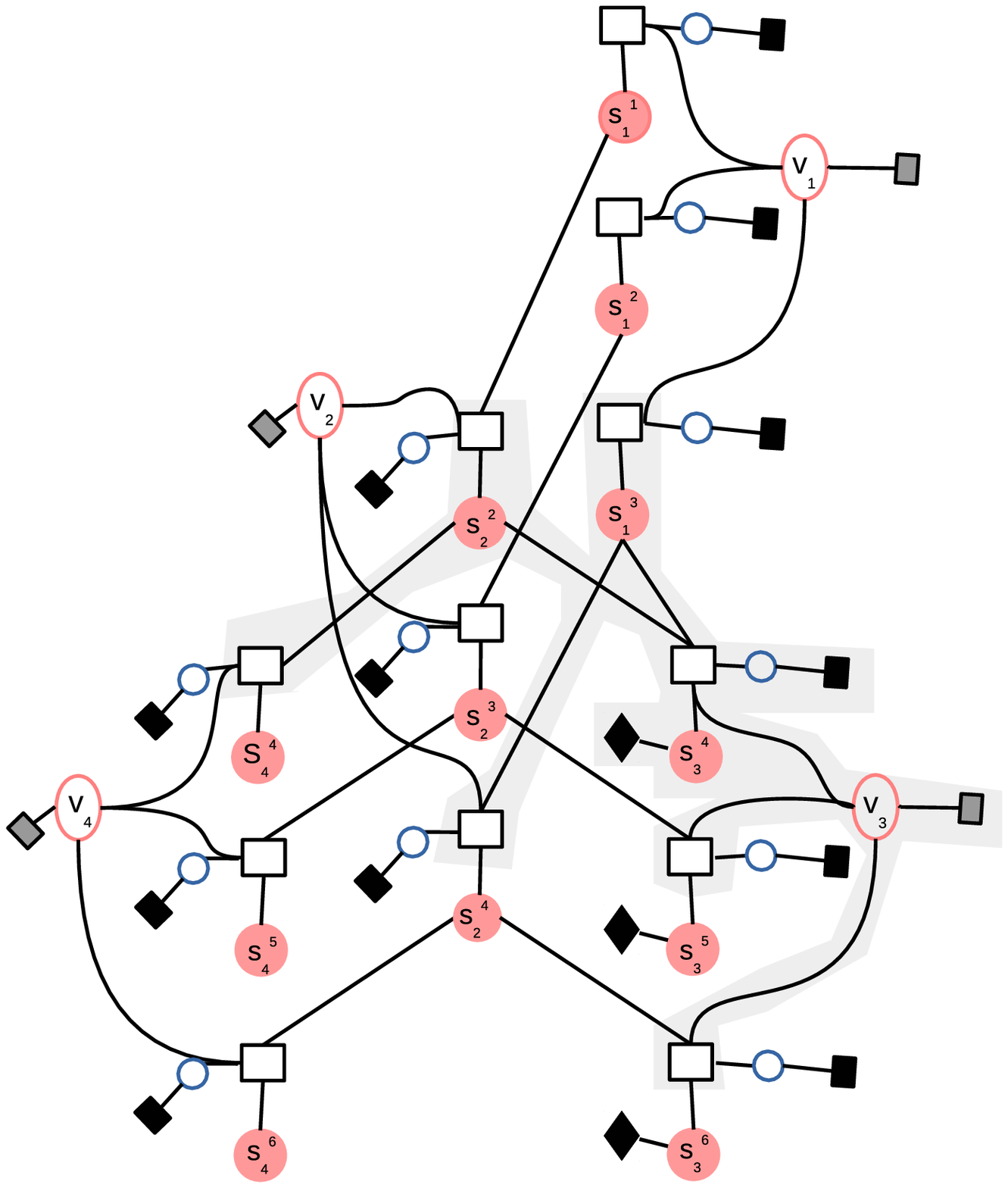}
 \caption{\label{fig:factorgraph}
 Factor graph representation for a small network. 
 Real graph of a example network. Circles are nodes of the network and the connection is directed an characterize by $\Delta t$ the delay time in appear contamination in node 2 once appeared in node 1.}  
% \caption{\label{extended graph}Extended graph of example network of figure \ref{fig:real graph} for a time period of analysis equal to $T$ steps. Circles are the state of a node in a time step and the connection is a direct relation between the state of two connected nodes.}
\end{figure}

A factor graph is a bipartite representation of the interactions between variables in a graphical model. In our case, we want to represent the probability 
\begin{equation}
P(S,V,Y | O) = \frac 1 Z e^{(-\beta\sum_{i,t}E_{i,t} +\sum_{i}\lambda \upsilon_i +\sum_{i,t}\gamma \yit +\sum_{o,t}\eta|s_o^t-{s^*}_o^t|)}.  \label{eq:Ptotal_ap}
\end{equation}
with two types of nodes, round nodes, representing variables ($\sit,\upsilon_i,\yit$) and angular nodes representing interactions. By interaction we mean every additive term in the exponent of the former probability distribution involving one or more variables. In Fig. \ref{fig:factorgraph} we show a representative part (for only 3 time intervals) of the factor graph of the small network of Fig. \ref{fig:time_extended}. Variables nodes are
\begin{itemize}
 \item full circles representing time extended nodes $s_i^t$,
 \item small empty circles representing $y_i^t$. The values of $i$ and $t$ (omitted) are those of the $s_i^t$ variable that is closest to them in the graph,
 \item empty ellipsoids correspond to $\upsilon_i$ variables.
\end{itemize}

The interactions between these variables are represented (in the same order they appear in eq. (\ref{eq:Ptotal_ap})) as 
\begin{itemize}
 \item empty squares corresponding to interactions $E_{i,t}(s_i^t,\upsilon_i, y_i^t, s_{\deltap_i}^{t-\Delta} )$ where $i,t$ correspond to the closest variable $s_i^t$,
 \item full gray squares corresponding to interactions $\lambda \upsilon_i$,
 \item full black squares corresponding to interactions $\gamma y_i^t$,
 \item full black rhomboids corresponding to the interaction with the observation $\eta|s_i^t-{s^*}_i^t|$ and is only present in the observed nodes.
\end{itemize}

Formally speaking, the construction of the factor graph, starting from the water distribution network graph and the observation given at some sensor nodes is outlined in algorithm \ref{alg:factor}. 

\begin{algorithm}
\caption{Factor Graph $G^*(V_v, V_f, E)$}
\label{alg:factor}
\begin{algorithmic}[1]
\REQUIRE Directed weighted graph $G(V,E,\Delta_T)$ for the water distribution network and sensed pattern $O(N,T)$.
\ENSURE Bipartite factor graph $G^*(V_v,V_f,E)$.
\STATE $V_v = \{\}, V_f=\{\}$.
\FOR {$(i,t) \in O$}
%  \STATE $V_i$ add node ($i,t$)
%  \STATE $V_f$ add node ($i,t$)
%  \STATE $V_v$ add node $n_i^t$; $V_f$ add node $a_i^t$;
%  \STATE $V_v \oplus= n_i^t$; $V_f \oplus= a_i^t$;
  \STATE $V_v = V_v \cup \{n_i^t\}$; $\quad V_f =  V_f \cup \{a_i^t\}$; \COMMENT{Add new nodes}
\ENDFOR
\FOR {$n_i^t \in V_v$}
 
   \FOR {$j \in \deltap_i$}
    \IF {$n_j^{t-\Delta_{ij}} \notin V_v$}
      \STATE $V_v = V_v \cup \{ n_j^{t-\Delta_{ij}}\}$; $\quad V_f= V_f \cup \{a_j^{t-\Delta_{ij}}\}$; \COMMENT{Add new nodes}
      \STATE $E =E\cup \{(a_i^t-n_j^{t-\Delta_{ij}})\}$ \COMMENT{Add edges between factor and variables nodes}
    \ENDIF
  \ENDFOR
\ENDFOR
\FOR {$n_i^t \in V_v$}
  \IF {$i>0$}
%    \STATE $V_v$ add node ($n_{-i}^{t}$)
%    \STATE $E$ add edge ($a_i^t-n_{-i}^t$)
    \STATE $V_v = V_v \cup \{n_{-i}^{t}\}$ \COMMENT{Add $y_i^t$ nodes}
    \STATE $E = E \cup \{(a_i^t-n_{-i}^t)\}$ \COMMENT{and connect to factor node}
    \IF {$n_i \notin V_v$}
      \STATE $V_v = V_v \cup \{n_i\}$ \COMMENT{Add $\upsilon_i$ nodes, if not yet present}
    \ENDIF
    \STATE $E = E \cup \{(a_i^t-n_i)\}$ \COMMENT{and connect to factor node}
  \ENDIF
\ENDFOR
\RETURN $G^*(V_v,V_f,E)$
\end{algorithmic}
\end{algorithm}

\subsection{Appendix 2: Belief Propagation}
\label{ap:bp}

Formally speaking, a consistent description of the full measure in terms of single variables and interacting variables distributions is achieved through a variational representation of the free energy $F = -\log Z$  of the problem in (\ref{eq:Ptotal}) as
\[  F[b_i(x_i),  b_{a}(x_{\partial a})] = \sum_a F_a[b(x_{\partial a})]  - \sum_i (d_i-1) F_i[b(x_i)]
\]
The free energy approximation sums the contribution of each factor node free energy, and subtracts the over-counting of the free energies of individual variables ($d_i$ is the number of factors that a variable interacts with). Each free energy term $F = U - T S$ consists, as usual, of an energetic part $U=\sum_{x_{\partial a}} b_a(x_{\partial a}) E_a(x_{\partial a})$ and $S = -\sum_{x_{\partial a}} b_a(x_{\partial a}) \log b_a(x_{\partial a})$, where the energetic term $E_a(\partial a)$ contains only the corresponding additive term in the exponential (\ref{eq:Ptotal_ap})

Minimization of this free energy over the distributions $b(\cdot)$ needs to respect the consistency among them: if two distributions share a common variable, the marginals should agree:
\begin{equation}
 \forall_{i,a\in \partial i} \:\: b_i(x_i) = \sum_{x \in a\setminus x_i} b_a(x_1,x_2,\ldots ) \label{eq:marginalization} 
\end{equation}
In order to achieve this, 
we introduce a set of Lagrange multipliers $m_{a \to i}(x_i)$ that are finally interpreted as messages flowing from every interaction towards every variable in it \cite{yedidia2005constructing}. The beliefs then are found in terms of these multipliers as in eq. (\ref{eq:bi}) 
% \begin{eqnarray}
%  b(s_i^t)& \propto \displaystyle \prod_{a \in \partial {\sit}} m_{a \to i}(\sit) \nonumber \\
%  b(y_i^t)& \propto  e^{-\gamma \yit} \displaystyle m_{a[i,t] \to y}(\yit)  \label{eq:bi-appendix} \\
%  b(\ui)& \propto  e^{-\lambda \ui} \displaystyle \prod_{a \in \partial {\ui}} m_{a \to i}(\ui)  \nonumber
% \end{eqnarray}
while the belief over every set of interacting variables in $ E_{i,t}$ are given as in eq. (\ref{eq:ba}) 

In algorithm \ref{alg:bp} we outline the main steps of the whole procedure presented in this paper, starting from the WDN graph up to the Belief Propagation fixed point iteration.

\begin{algorithm}
\caption{Belief propagation for contamination source detection}
\label{alg:bp}
\begin{algorithmic}[1]
\REQUIRE

\begin{itemize}
 \item $G$ directed weighted graph for the water distribution network.
 \item $O$ patterns observed at sensor nodes.
 \item $\epsilon$ fixed point goal precision.
 \item $N_{\tiny{iter}}$ maximum number of fixed point iterations.
 \item $\lambda$, $\gamma$, $\eta$ y $\beta$  parameters for the probabilistic model.
\end{itemize}
\ENSURE Marginal probabilities over each binary variable $\forall_i b(\ui)$,  $\forall_{[i,t]} b(\yit)$ y $ \forall_{[i,t]} b(\sit)$.
%\STATE Se lee el grafo y se crean los estados $\sit$, $\ui$, $\yit$, los nodos factores $f_i^t$ de cada nodo y se crean los conjuntos de estados $s_j^t \in \partial_i^+$ padres y hijos $s_k^t \in \partial_i^-$ de cada estado $\sit$.
\STATE $G' = (V_v,V_f,E) = \mbox{FactorGraph}(G,T)$
\FOR{ $i \in V_v \cup V_f $}
  \STATE $b_i(+1) = b_i(0) = 0.5$
  \FOR{ $a \in \partial i$}
  \STATE $m_{a\to i}(+1) =m_{a\to i}(0)= 0.5$
  \ENDFOR
\ENDFOR
%\STATE Se inicializan los mensages de cada estado a 0.5.
%\STATE Se lee el archivo del sensor y se inicializan los mensajes de cada estado con los par\'ametros.
\WHILE{ $\mbox{maxdiff}>\epsilon$ and $it < N_{\tiny{iters}} $}
\STATE $\mbox{maxdiff} = 0.0$
\FOR{ $i \in \mbox{Nodes}(G') $}
  \FOR{ $a \in \partial i$}
  \STATE $m'_{a\to i}(1) = m_{a\to i} (1)$ \COMMENT{Make a copy to check how much it changed}
  \STATE Update $m_{a\to i}$ using eq. (\ref{eq:mensaje}), and normalize $m_{a\to i}$.
  \STATE $\mbox{maxdiff} = \max(\mbox{maxdiff}, |m_{a\to i}(1) - m'_{a\to i}(1)| )$ \COMMENT{Keep maximum change in messages }
  \ENDFOR
\ENDFOR
\ENDWHILE
%\WHILE {Los mensajes de los estados var\'ien m\'as que un valor $\varepsilon=0.000001$}
%\STATE Actualiza los mensajes de todos los nodos del sistema usando la ecuaci\'on \ref{eq:mensaje}
%\ENDWHILE
\IF{ $\mbox{maxdiff}<\epsilon$}
\RETURN $b(s_i^t)$, $b(\yit)$ y $b(\ui)$. Report as probable origin of contamination every node $i$ such that $b_i(\ui=1) > 0.5$.
\ELSE
\RETURN Not Converged.
\ENDIF
\end{algorithmic}
\end{algorithm}

% \label{section:references}
\bibliographystyle{unsrt}

\bibliography{bibliografia1}

\begin{thebibliography}{10}

\bibitem{REVERSE}
E.~Salomons and A.~Ostfeld.
\newblock Identification of possible contamination sources using reverse
  hydraulic simulation.
\newblock In {\em 12th Annual International Symposium on Water Distribution
  Systems Analysis, Tucson, Arizona, USA, published on CD}, 2010.

\bibitem{laird2006mixed}
Carl~D Laird, Lorenz~T Biegler, and Bart~G van Bloemen~Waanders.
\newblock Mixed-integer approach for obtaining unique solutions in source
  inversion of water networks.
\newblock {\em Journal of Water Resources Planning and Management},
  132(4):242--251, 2006.

\bibitem{laird2007real}
Carl~D Laird, Lorenz~T Biegler, and Bart~G van Bloemen~Waanders.
\newblock Real-time, large-scale optimization of water network systems using a
  subdomain approach.
\newblock In {\em Real-Time PDE-Constrained Optimization}, pages 289--306.
  SIAM, 2007.

\bibitem{liu2011logistic}
Li~Liu, A~Sankarasubramanian, and S~Ranji Ranjithan.
\newblock Logistic regression analysis to estimate contaminant sources in water
  distribution systems.
\newblock {\em Journal of Hydroinformatics}, 13(3):545--557, 2011.

\bibitem{preis2006contamination}
A.~Preis and A.~Ostfeld.
\newblock Contamination source identification in water systems: A hybrid model
  trees--linear programming scheme.
\newblock {\em Jour.Wat.Res.Plan.Manag.}, 132(4):263--273, 2006.

\bibitem{huang2009data}
J.~J. Huang and E.~A. McBean.
\newblock Data mining to identify contaminant event locations in water
  distribution systems.
\newblock {\em Jour.Wat.Res.Plan.Manag.}, 135(6):466--474, 2009.

\bibitem{hu2015mapreduce}
Ch. Hu, J.~Zhao, X.~Yan, D.~Zeng, and S.~Guo.
\newblock A mapreduce based parallel niche genetic algorithm for contaminant
  source identification in water distribution network.
\newblock {\em Ad Hoc Net.}, 35:116--126, 2015.

\bibitem{guan}
J.~Guan, M.~M. Aral, M.~L. Maslia, W.~M., and Grayman.
\newblock Identification of contaminant sources in water distribution systems
  using simulation--optimization method: case study.
\newblock {\em Jour.Wat.Res.Plan.Manag.}, 132(4):252--262, 2006.

\bibitem{tao2012identification}
T.~Tao, X.~Fu Y-j. Lu, and K~l.~Xin.
\newblock Identification of sources of pollution and contamination in water
  distribution networks based on pattern recognition.
\newblock {\em Jour.Zhej.Univ-Scie.A}, 13(7):559--570, 2012.

\bibitem{kumar2012contaminant}
Jitendra Kumar, E~Downey Brill, G~Mahinthakumar, and S~Ranji Ranjithan.
\newblock Contaminant source characterization in water distribution systems
  using binary signals.
\newblock {\em Journal of Hydroinformatics}, 14(3):585--602, 2012.

\bibitem{khancontamination}
Md~Aminul~Islam Khan and Bijit~Kumar Banik.
\newblock Contamination source characterization in water distribution network.
\newblock {\em Global Science and Technology Journal}, 5(1):44--55, 2017.

\bibitem{wang2011bayesian}
H.~Wang and K.~W. Harrison.
\newblock Bayesian update method for contaminant source characterization in
  water distribution systems.
\newblock {\em Jour.Wat.Res.Plan.Manag.}, 139(1):13--22, 2011.

\bibitem{wang2013bayesian}
H.~Wang and K.~W. Harrison.
\newblock Bayesian approach to contaminant source characterization in water
  distribution systems: adaptive sampling framework.
\newblock {\em Stoch.Env.Res.Risk.Assess.}, 27(8):1921--1928, 2013.

\bibitem{wang2012improving}
Hui Wang and Kenneth~W Harrison.
\newblock Improving efficiency of the bayesian approach to water distribution
  contaminant source characterization with support vector regression.
\newblock {\em Journal of Water Resources Planning and Management},
  140(1):3--11, 2012.

\bibitem{propato}
M.~Propato, F.~Sarrazy, and M.~Tryby.
\newblock Linear algebra and minimum relative entropy to investigate
  contamination events in drinking water systems.
\newblock {\em Jour.Wat.Res.Plan.Manag.}, 136(4):483--492, 2009.

\bibitem{barandouzi2016probabilistic}
Mehdy Barandouzi and Reza Kerachian.
\newblock Probabilistic contaminant source identification in water distribution
  infrastructure systems.
\newblock {\em Civil Engineering Infrastructures Journal}, 49(2):311--326,
  2016.

\bibitem{cristo2008pollution}
C.~D. Cristo and A.~Leopardi.
\newblock Pollution source identification of accidental contamination in water
  distribution networks.
\newblock {\em Jour.Wat.Res.Plan.Manag.}, 134(2):197--202, 2008.

\bibitem{perelman2010bayesian}
A.~L.~Perelman and Ostfeld.
\newblock Bayesian networks for estimating contaminant source and propagation
  in a water distribution system using cluster structure.
\newblock In {\em Water Distribution Systems Analysis 2010}, pages 426--435.
  2010.

\bibitem{paper}
Alfredo Braunstein, Alejandro Lage-Castellanos, and Ernesto Ortega.
\newblock Contamination source inference in water distribution networks.
\newblock {\em Rev.Cub.Fís}, 34(2):100--107, 2017.

\bibitem{lokhov2014inferring}
H.~Ohta A.~Y.~Lokhov, M.~M{\'e}zard and L.~Zdeborov{\'a}.
\newblock Inferring the origin of an epidemic with a dynamic message-passing
  algorithm.
\newblock {\em Phy.Rev.E}, 90(1):012801, 2014.

\bibitem{luo2014identify}
W.~P~Tay W.~Luo and M.~Leng.
\newblock How to identify an infection source with limited observations.
\newblock {\em IEEE Jour.Sel.Top.Sign.Proc.}, 8(4):586--597, 2014.

\bibitem{zhu2016information}
K.~Zhu and L.~Ying.
\newblock Information source detection in the sir model: A sample-path-based
  approach.
\newblock {\em IEEE/ACM Trans.Net.}, 24(1):408--421, 2016.

\bibitem{Ale}
F.~Altarelli, A.~Braunstein, L.~Dall’Asta, A.~Lage-Castellanos, and
  R.~Zecchina.
\newblock Bayesian inference of epidemics on networks via belief propagation.
\newblock {\em Phy.Rev.Lett.}, 112(11):118701, 2014.

\bibitem{Todini}
E.~Todini and S.~Pilati.
\newblock A gradient algorithm for the analysis of pipe networks.
\newblock In {\em Computer applications in water supply: vol. 1---systems
  analysis and simulation}, pages 1--20. Research Studies Press Ltd., 1988.

\bibitem{yedidia2005constructing}
Jonathan~S Yedidia, William~T Freeman, and Yair Weiss.
\newblock Constructing free-energy approximations and generalized belief
  propagation algorithms.
\newblock {\em IEEE Transactions on information theory}, 51(7):2282--2312,
  2005.

\bibitem{Modena}
Large problems.
\newblock Center of Water Systems at University of Exeter.

\end{thebibliography}

%\bibliography{ascexmpl-new}
%

\end{document}